\documentclass{article}

\usepackage{graphicx}  
\usepackage{amsmath}   
\usepackage{amssymb}   

\usepackage[active]{srcltx}

\def\eq#1{{Eq.~(\ref{#1})}}

\usepackage{bm}
\usepackage{latexsym}
\usepackage{dcolumn}
\usepackage{amsmath,amsfonts,amssymb}
\usepackage{graphicx,epsfig}
\usepackage{color}
\usepackage{enumitem}
\usepackage[active]{srcltx}

\def \gm {q}
\def \rmod {{\mathcal{R}}}

\def \lp { L_P }

\def\nn{\nonumber}

\def\l{\left}
\def\r{\right}
\def\DM{\mathrm{d}}

\def\cc{cosmological\ constant}
\def\bdot{\bm{\dot}}

\def\bddot{\bm{\ddot}}
\reversemarginpar

\def \mA {A\l[\sigma;\lp \r]}

  \title{Grin of the Cheshire cat:  Entropy density of spacetime  as a relic from quantum gravity}
\author{Dawood Kothawala\footnote{dawood@physics.iitm.ac.in}\\
Department of Physics, IIT Madras, Chennai - 600 036, India.\\
T. Padmanabhan\footnote{email: paddy@iucaa.ernet.in}, \\
IUCAA,
 Post Bag 4, Ganeshkhind, Pune - 411 007, India 
}

\begin{document}
  
  \maketitle
  
  \begin{abstract}
\noindent 
There is considerable amount of evidence to suggest that the field equations of gravity have the same status as, say, the equations describing an emergent phenomenon like elasticity. In fact, it is possible to derive the field equations from a thermodynamic variational principle in which a set of normalized vector fields are varied rather than the metric. We show that this variational principle can arise as a low energy ($L_P = (G\hbar/c^3)^{1/2} \to 0$) relic of a plausible non-perturbative effect of quantum gravity, viz. the existence of a zero-point-length in the spacetime. Our result is non-perturbative in the following sense: If we modify the geodesic distance in a spacetime by introducing a zero-point-length, to incorporate some effects of quantum gravity, and take the limit $L_P  \to 0$ of the Ricci scalar of the modified metric, we end up getting a nontrivial, leading order ($L_P$ - independent) term.  
\textit{This term  is identical to the expression for entropy density of spacetime used previously in the emergent gravity approach.}
This reconfirms the idea that the microscopic degrees of freedom of the spacetime, when properly described in the full theory, could lead to an effective description of geometry in terms of a thermodynamic variational principle. This is  conceptually similar to the emergence of thermodynamics from the mechanics of, say, molecules. The approach also has important implications for the cosmological constant which are briefly discussed. \end{abstract}

 \section{Introduction, Motivation and Summary}
 
Recent work has gone a long way in demonstrating that gravitational field equations can be thought of as having the same conceptual status as equations in emergent phenomena like fluid dynamics or elasticity (\cite{tpreviews,grtp}). Two strongest pieces of evidence, amongst many, which support this point of view are the following: 

\begin{itemize}

 \item It is possible to obtain \cite{entropy-functional,llreview} the field equations of a large class of gravitational theories including, but not limited to, Einstein's theory from an alternative, thermodynamic variational principle. In this approach, one starts with a well defined thermodynamic potential $\mathcal{S}$[$\nabla n, n$] which depends on a  vector field $n^i$ of constant norm and has \cite{grtp} the interpretation as the gravitational heat density of spacetime:
\begin{equation}
 \mathcal{S}\propto [(\nabla_in^i)^2-\nabla_i n^j \nabla_jn^i]=R_{ab} n^an^b +(\text{total divergence})
\label{ts}
\end{equation}

 Extremising $\mathcal{S}$ with respect to all vector fields $n^i$ simultaneously, leads to a constraint on the background metric which turns out to be identical to the field equations. (Adding the appropriate matter heat density and extremising the total heat density will lead to field equations with source; in this paper, we will be mainly concerned about pure gravity.)

\item The time evolution of the spacetime geometry can be described through an equation which is mathematically equivalent to, say, Einstein's equation but can be written and interpreted entirely in terms of surface ($N_{\rm sur}$) and bulk ($N_{\rm bulk}$) degrees of freedom \cite{grtp,tpsur}. When the metric is independent of time \cite{tpsur} we get, $N_{\rm sur} = N_{\rm bulk}$ (``holographic equipartition'') and --- in the most general context --- the time evolution of the metric is driven by \cite{grtp} the difference ($N_{\rm sur} - N_{\rm bulk}$). Thus, not only the variational principle \textit{but even the resulting field equation} can be expressed in a thermodynamic language that brings to the fore the importance of microscopic degrees of freedom in the bulk and in the boundary.

\end{itemize}

These results suggest that one should interpret the standard gravitational dynamics as the thermodynamic limit of some underlying statistical physics which deals with the (as yet unknown) atoms of spacetime. Eventually, when we discover the fundamental laws governing the dynamics of these microscopic degrees of freedom, we will \textit{also} discover and understand the limiting process by which the thermodynamic variational principle, based on $\mathcal{S}$, can be obtained. While this possibility sounds natural, there are couple of puzzling features related to this issue. 

First, it seems reasonable to assume that classical equations of gravity should arise in the $L_P \to 0$ limit of the full quantum dynamics. In that case, one would (naively, as it turns out) expect the leading order action functional of classical gravity to be the standard Einstein-Hilbert action with higher order ($L_P$ dependent) corrections. \textit{But, this does not seem to be the direction which is suggested by the emergent gravity paradigm.} There are strong conceptual reasons to believe (including the problem of the cosmological constant) that the metric should \textit{not} be the dynamical variable which is varied in the low energy effective action of the theory and \textit{ipso facto} Einstein-Hilbert action will not be the correct, $L_P\to 0$ limit, of the microscopic variational functional. 
Instead, the emergent gravity paradigm suggests using $\mathcal{S} (\nabla n, n)$  and varying $n^i$. It is not clear how this transmutation in the variational principle (from $R$ to $R_{ab} n^a n^b$) arises.

Let us elaborate on this point which suggests a radical departure from the conventional wisdom. The usual approaches to quantum gravity presuppose that the metric will \textit{continue} to be a dynamical degree of freedom  in the quantum gravitational domain (in some form or the other) \textit{because} it is the dynamical variable in the classical limit. But  the emergent gravity paradigm suggests something different. In the classical limit, it does \textit{not} treat the usual metric as the degree of freedom to be varied in an extremum principle and hence there is \textit{no reason to expect this metric to play a direct role in the quantum theory either}. If Einstein's equations are like equations of elasticity, treating the metric as a quantum variable is like quantizing elasticity; we will then get gravitons as analogues of phonons but the real microscopic degrees of freedom will be quite different --- in a solid or in a spacetime.

Second, the variational principle based on $\mathcal{S}$ uses a vector field of constant norm. It is not clear, a priori, how a quantum gravitational variational principle will produce such a vector field which survives in the low energy limit of $L_P\to 0$. Where does it come from and why should it be varied in the low energy limit rather than the metric (which is the conventional point of view) are the questions that need to be answered in a microscopic theory.

We will answer these questions in this paper. 

We will show that there exists a natural variational principle which could have a microscopic origin and can lead to the variational principle based on $\mathcal{S}$ in the $L_P\to 0$ limit. As we will see, the limiting process  introduces in a natural fashion a normalized vector field $n^i$ and  the field equations of the low energy theory can be obtained by varying  this vector field. This demonstrates the possibility that the thermodynamic variational principle can indeed have a microscopic origin. We will see that the limiting process is subtle and mathematically non-trivial. It is this non-triviality which leads to a leading order ($L_P\to 0$) variational principle that is quite different  from the Einstein-Hilbert action.

The key new idea is to work with the biscalar $\sigma^2 (p,P)$ which is the geodesic distance between two events $p$ and $P$ in any spacetime (with $\Omega\equiv(1/2)\sigma^2$ being the so called Synge world function). Locally, this function is related to the metric by the usual Hamilton-Jacobi equation, $(1/2)g^{ab}\nabla_a \Omega \nabla_b \Omega=\Omega$, which takes the form:
\begin{equation}
g^{ab}(x) \nabla_a \sigma^2(x,x') \; \nabla_b \sigma^2(x,x')= 4 \sigma^2(x,x') 
\label{HJ1}
\end{equation} 

Indeed,  all the information about the spacetime geometry can be shown to be encoded in $\sigma^2$. This is because the metric can be obtained \cite{poisson-lrr} from the coincidence limit  of covariant derivatives of $\sigma^2$ by: 
\begin{equation}
 g_{ab} = \lim_{x'\to x} \l[{\nabla}_a {\nabla}_b \Omega(x,x') \r] 
\label{gR} 
\end{equation}
and, of course, all other geometrical properties can be obtained from the metric.
In other words, \textit{all} of classical gravity and all of spacetime dynamics can be described entirely in terms of the biscalar function $\sigma^2 (p,P)$. Therefore, one can trade off the (local) degrees of freedom associated with the metric for the degrees of freedom represented by the (nonlocal) object $\sigma^2 (p,P)$.

The major advantage of using the geodesic distance instead of the metric is the following. We have no clue how quantum gravitational effects modify the notion of the metric at short distances and what kind of effective description is called for in the quasi-classical domain. However, there is significant amount of evidence \cite{zpl,zpltp}  to suggest that quantum gravity introduces a zero-point-length to the spacetime in the sense of:
\begin{equation}
 \lim_{p\to P}\ \langle \sigma^2(p,P) \rangle = \lp^2
\label{zpl1}
\end{equation} 
where
the $\langle \cdots \rangle$ denotes averaging over metric fluctuations and $\lp$ is a fundamental length scale,  \textit{of the order of} Planck length ($\approx 10^{-33}$ cm). 
This suggests that 
one can capture the lowest order quantum gravitational effects by introducing a zero-point-length to the spacetime by modifying 
\begin{equation}
 \sigma^2 \to \sigma^2 + \lp^2.
\label{funda}
\end{equation}
For example, this changes the coincidence limit of Green's functions and serves as a Lorentz-invariant UV regulator. The  Euclidean propagator for the massless scalar field, for example, gets replaced by:
\begin{equation}
G(\sigma^2)\propto (\sigma^2)^{-(D-2)/2}\to (\sigma^2+L_P^2)^{-(D-2)/2}
\label{modG}
\end{equation} 
near the coincidence limit \cite{zpl} (and the same holds for the massive propagator as well). The logic and justification for such a result have been presented in several previous papers \cite{zpl,zpltp,pid} and will not be repeated here. But the key point to note is that the correction to $\sigma^2$ in \eq{funda} is \textit{universal} and captures some basic features of quantum gravity. 
Equations (\ref{zpl1}) and (\ref{funda}) should be considered as purely non-perturbative results in quantum gravity arising from the quantum gravitational averaging of $\sigma^2 (p,P |g_{ab})$ over the fluctuations of the metric $g_{ab}$.  

 We have, of course, no clear idea as to how to describe the spacetime at Planck scales where the notions of differential manifold, metric etc. may  break down completely. On the other hand, these features emerge in the long-wavelength limit, at scales much larger than $L_P$. It seems reasonable to assume that there could exist an in-between, quasi-classical domain,  that interfaces classical and quantum gravity, in which we can still talk about a metric, spacetime interval etc but with some modifications arising from the quantum gravity. The results in \eq{zpl1} and \eq{funda} motivate one to proceed along the following lines to describe physics in the quasi-classical domain:

\begin{enumerate}
 \item Suppose we are interested in a classical spacetime with a given metric $g_{ab}$ or, equivalently, a given biscalar $\sigma^2 (p,P) $ corresponding to that metric. Assume for a moment that we can find another second rank, symmetric, bitensor  $\gm_{ab}(p, P; L_P^2)$ (which we will call `qmetric' for reasons which will be clear soon) such that its geodesic distance is $\sigma^2 (p,P) + L_P^2$.
\label{page:ricci}
\item Since it can be argued that $ \sigma^2 \to \sigma^2 + L_P^2$ captures the  quantum gravitational effects in the quasi-classical domain, it seems reasonable to think of a variational principle based on  $\gm_{ab}(p,P;L_P^2 )$ to achieve the same.  This, in turn, motivates us to consider the Einstein-Hilbert action functional for  $\gm_{ab}(p, P; L_P^2)$ which is constructed from the Ricci biscalar for the qmetric $\gm_{ab}(p, P; L_P^2)$ using the standard formula which connects  Ricci scalar to the metric.

\item The qmetric is, of course, a non-local  bitensor depending on \textit{two} events $p$ and $P$ (and on $L_P^2$). Hence the Ricci biscalar $\rmod(p,P;L_P^2)$ obtained from it will also depend on two events $p$ and $P$ and on $L_P^2$. Obviously, we need to take a suitable limit of $p\to P$ as well as $L_P \to 0$ to obtain a local expression analogous to the Ricci scalar for the geometry. The resulting scalar 
\begin{equation}
L_{eff} \equiv  \lim_{\lp\to0}\lim_{p\to P} \rmod(p,P;\lp^2) 
\end{equation} 
can then be interpreted as the correct, low-energy, functional to be used in the variational principle.

\end{enumerate}

Incredibly enough, strange --- but nice --- things happen along the way when we attempt to carry out the logical steps of the above program. 

To begin with, the qmetric is not a metric in the standard sense of differential geometry. This should be obvious because any geodesic interval obtained from a genuine metric, by integrating along a geodesic from event $p$ to event $P$, is guaranteed to vanish when we take the coincidence limit of $p\to P$.  We are never going to get the zero-point-length modification of the geodesic interval in \eq{funda} if the qmetric is a genuine, local, metric and the limiting process is non-singular. As we shall show, the qmetric depends explicitly on both $p$ and $P$ (making it a \textit{nonlocal} bitensor) and has a singular structure in the coincidence limit ---  which allows the geodesic interval constructed from the qmetric to be non-vanishing  in the coincidence limit. The fact that this result arises in a very natural fashion is noteworthy. 

Second --- and  the most surprising feature --- is the form of the coincidence limit of the Ricci biscalar. It turns out that when the coincidence limit $p\to P$ is taken in the Ricci biscalar for the qmetric, the leading term is precisely $\mathcal{S}$ given in \eq{ts} and \textit{not} the Ricci scalar for the original metric! That is, we get:
\begin{equation}
L_{eff} \equiv  \lim_{\lp\to 0}\lim_{p\to P} \rmod (p,P;\lp^2) \propto  \mathcal{S}(P) 
\label{explim}                                                                                                                                                                                                                                                                                                                                                                                                                                                                                  
\end{equation}
                                                                                                                                                                                                                                                                                                                                                                                                                                                                            In other words, \textit{starting from the Ricci biscalar for the qmetric, taking the $p\to P$ limit  and then taking $L_P \to 0$ limit does not lead to the Ricci scalar for the original metric} but to $\mathcal{S}$ used in emergent gravity paradigm! The final result depends on an extra vector field of constant norm arising from the derivatives of the geodesic interval. 

The entire process  involves taking the limits in a particular order so that the local quantities are defined from the non-local entities in a specific manner. Such operations are well justified by physical considerations at this stage and will probably acquire a firmer mathematical basis when we understand quantum gravity better. 
What is obvious from our result --- and is quite significant --- is the following conceptual fact: Quantum gravitational structures which depend on $L_P$ can lead to unexpected semi-classical relics when 
$L_P \to 0$ limit is taken, if quantum gravity is non-analytic in $L_P$. There is considerable amount of evidence that this could indeed be the case. 

An analogy may be useful to clarify this point. We know that the classical theory of elasticity should be obtainable from the quantum dynamics of a solid by taking an appropriate continuum limit. One would have thought that such a classical limit of a continuum solid can be obtained by taking the  $\hbar \to 0$ limit of  the quantum theory. This is however too naive. If we take the (mathematically) strict $\hbar \to 0$ limit of a quantum solid, each atom will collapse to a singularity because atoms cannot exist in the  $\hbar \to 0$ limit.  To get the proper limit, we have to keep ``the $\hbar$ inside the atoms'' to be non-zero (ensuring the existence of atoms) and let ``all other $\hbar$'s'' to vanish. The real classical world cannot be described as a sum of a leading order, $\hbar $-independent, phenomenon with $\hbar $-dependent corrections. Figuratively speaking, classical world is non-analytic in $\hbar$ because matter made of atoms cannot exist in the strict $\hbar\to0$ limit.

In a similar manner, one could argue that $\hbar \to 0$ limit is non-trivial in proceeding from quantum to classical limit of spacetime. There is fair amount of evidence that what is fundamental in quantum gravity is the quantum of area, $L_P^2\propto\mathcal{A}_P = (G\hbar / c^3)$. If we write the Newton's law of gravitation in two equivalent forms as 
\begin{equation}
 F = G \frac{m_1m_2}{r^2} = \frac{ \mathcal{A}_P c^3}{\hbar} \frac{m_1m_2}{r^2}
\end{equation}  
it is clear that the limit $\hbar \to 0$ can have widely different behaviour depending on which form we use. Conventionally, one thinks of $G$ as independent of $\hbar$ and the first form of the equation is oblivious to the process of taking $\hbar \to 0$. But if $G$ is an \textit{emergent} constant, like Young's modulus or conductivity of a solid, then it can have a non-trivial, implicit, dependence on $\hbar$. If we further assume, based on \eq{funda} that $\mathcal{A}_P$ is the quantity which is independent of $\hbar$, and characterizes quantum spacetime, then a completely different picture emerges \cite{tpgrf02}. Now we need to use the second form of the equation to study, say, the planetary motion. The $\hbar \to 0$ limit does not exist and Newton's law of gravitation is non-analytic in $\hbar$; there is no such thing as classical gravity just as there is no such thing as classical solid since neither could exist in the strictly $\hbar \to 0$ limit. 

Most of our conventional intuition about quantum gravity is built on the idea that classical gravity can be obtained as a ``Taylor series expansion'' in $L_P^2$ starting from quantum gravity. This, in turn, assumes that all quantum gravitational effects are analytic in $L_P^2$ and will lead to some sensible classical limits when $L_P \to 0$ limit is taken. Some thought shows that \textit{this is a highly questionable assumption} and we should take seriously the possibility that quantum gravity could have features which are non-analytic in $L_P$. In that case, the process of taking  limits  will involve manipulating singular quantities leading to unexpected (but interesting) results.  The emergent nature of gravitational field equations can arise from such a non-trivial limiting process which we illustrate here.

There is a different conceptual bonus from this analysis. There are, arguably, three areas of contact and conflict between the principles of quantum theory and gravity:
	(i) Thermodynamics of spacetime horizons.
	(ii) The singularity problem in classical gravity, especially in cosmology and black hole physics.
	(iii) The problem of the cosmological constant.
There have been repeated suggestions in the literature that quantum gravity will have something non-trivial to say about all these three issues. It is already known that emergent gravity paradigm provides deep insights into items (i) and (iii) but has been silent about the singularity problem. On the other hand, the prescription in \eq{funda} has the potentiality of tackling the singularity problem by making the coincidence limit finite due to the fluctuations of the metric. We show here that the prescription in \eq{funda} --- which has been discussed in the past in connection with the singularity resolution and as a UV-regulator ---  can also lead to the variational principle in emergent gravity. So the present work provides a unifying thread linking the three issues listed above which is conceptually rather pleasing.

We shall now describe some of the mathematical details of this procedure. Rest of the sections of the paper are organized as follows. In Sec. \ref{sec:rs}, we present the exact form of the Ricci biscalar for the qmetric and obtain from it a local scalar which is the natural candidate for the quantum corrected Ricci scalar. We then compare this object with the Ricci scalar of the original spacetime, which reveals the non-triviality of the $L_P \to 0$ limit. In Sec. \ref{sec:implications}, we analyze our result in the context of emergent gravity paradigm, and discuss how various pieces fit nicely to demonstrate the naturalness and inevitability of this paradigm. In particular, we discuss how our results strongly suggest that the \cc\;might be related to a non-local relic of the small scale structure of spacetime. Finally, in Sec. \ref{sec:outlook}, we discuss some important conceptual points which deserve a deeper investigation.

{\it \textbf{Notation}}: We work in $D$ dimensions, and use the sign convention $(-, +, +, \ldots)$ for Lorentzian spaces. \textit{Most of our analysis is best viewed as done in a Euclidean space, with analytic continuation done right in the end.} This is implicit in the calculations even when we do not state in explicitly. In this sense, the parameter $\epsilon$ which is the norm of the normalized tangent vector $n^i$ to geodesics is simply $+1$. We have nevertheless kept it in the equations to help keep track of sign issues arising for timelike geodesics after analytic continuation.

\section{The qmetric and the corresponding Ricci biscalar} \label{sec:rs} 
We will now carry out the steps described in items (1) to (3) in page \pageref{page:ricci}. Our starting point is the result derived in \cite{dk-ml}.
This result allows us to associate a second rank, symmetric, bitensor $\gm_{ab}(p,P)$ with a spacetime having the metric $g_{ab}$ and the geodesic distance $\sigma^2(p,P)$ such that the following two properties are satisfied: 
(i) The modified geodesic distance 
\begin{equation}
\sigma_{(q)} {}^2 = \sigma^2 + \lp^2
\label{eq:final1}
\end{equation}
(where $\sigma_{(q)} {}^2 \equiv \sigma^2(p,P \vert \gm_{ab})$ and $\sigma^2 \equiv \sigma^2(p,P \vert g_{ab})$ for simplifying the  notation) satisfies the analogue of \eq{HJ1} with the metric replaced by the qmetric. That is:
\begin{equation}
\gm^{ab} \nabla_a \sigma_{(q)} {}^2 \; \nabla_b \sigma_{(q)} {}^2= 4 \sigma_{(q)} {}^2 
\label{HJ2}
\end{equation}
(ii) The Euclidean propagator for the massless scalar field, with the modification in \eq{modG}, satisfies the usual Green function equation $\sqrt{-g} \; \square G=\delta(x,x')$ with the qmetric replacing the metric in the operator on the left-hand-side. These two conditions allow us to determine the form of the qmetric to be:

\begin{equation}
\gm_{ab}(p,P; \lp^2) = \mA \; g_{ab}(p) -  \epsilon \l( \mA  - \frac{1}{\mA} \r) \; n_a(p;P) \; n_b(p;P)
\label{eq:key1}
\end{equation}
with
\begin{equation}
\mA = 1 + \frac{\lp^2}{\sigma^2}; \qquad n_a = \frac{\nabla_a \sigma^2}{2 \sqrt{\epsilon \sigma^2}}
\label{n1}
\end{equation}
In fact, it is possible to generalize the notion of qmetric for an \textit{arbitrary} modification of the geodesic interval $\sigma^2\to S(\sigma^2)$; an outline of derivation for this general case is given in Appendix \ref{app:generalized-qmetric} which covers the above result as a special case.
Derivation of the above result, for the modification of the geodesic interval in \eq{eq:final1}, and its  implications (such as the effect on short-distance behavior of Green's functions and on spacetime singularities) are discussed in ref.\cite{dk-ml}.

One can think of \eq{eq:key1} as associating a qmetric with every metric such that the geodesic distance computed from the qmetric is $(\sigma^2 + \lp^2)$ if the geodesic distance computed from the original metric is $\sigma^2$.
 Throughout the paper we think of $P$ as a fixed `base event' and $p$ as the variable `field event'. All derivatives, covariant or partial, are taken at the spacetime event $p$. We see that when the geodesic distance between the two events is much larger than the Planck length ($\sigma^2(p,P)/L_P^2\gg 1$), the qmetric reduces to the background metric $\gm_{ab}(p,P; \lp^2)\to g_{ab}(p)$. In this sense we can think of qmetric as describing the nonlocal, quantum gravitational effects near the Planck scale, arising from the existence of a zero-point-length.

We will now attempt to relate the curvature bi-invariants (obtained by the usual formulas treating it like a metric) of the  qmetric $\bm \gm$   with those of the original metric 
$\bm g$. Needless to say, this is a formidable task, since the  qmetric is a sum of two terms: (i) a piece which is conformal to $\bm g$, and (ii) a piece depending on the tangent vector $\bm n$ connecting $P$ to $p$. This complicates the evaluation of the full curvature tensor of $\bm \gm$  in terms of that of $\bm g$. Fortunately, for the purpose of this paper,   we  only need the Ricci biscalar corresponding to the qmetric which can be derived by  a few tricks involving the Gauss-Codazzi relation.\footnote{This analysis also sheds some light on more general aspects of intrinsic and extrinsic geometries of metrics related in a manner similar to Eq.~(\ref{eq:key1}), which might be relevant in their own right from a purely differential geometric point of view; these will be presented in a separate paper \cite{WIP}.} 

To begin with, note that the deformation is characterized by the function $\mA$ which is only a function(al) of $\sigma^2$. Further, as was shown in \cite{dk-ml}, the {\it induced} geometry on $\sigma^2=$const surface undergoes a {\it conformal} deformation
$
h_{ab} \to \mA \; h_{ab}
$
which leads to a rather simple relation for the {\it induced} Ricci scalars which appear in the Gauss-Codazzi equations. The two facts above suggest that it might be mathematically convenient to consider the {\it foliation defined by $\sigma^2=$constant surfaces}.
The extrinsic curvature for this foliation can be derived using the identities satisfied by the geodesic distance function  which have been summarized, for example, in \cite{christenson-worldf}.
Using this procedure we can compute the explicit form of the Ricci scalar for the qmetric. The outline of the steps for the general case is given in Appendix \ref{app:rs-eval-details} and the explicit derivation of the final result for the special case are given in Appendix \ref{app:example}. 

Thus, after some lengthy but straightforward algebra, the final expression is found to be
\begin{equation}
\rmod(p;P,\lp) = 
R(p) 
\; - \; 
\l(1 - \frac{1}{A} \r) \mathcal Y \; - \; \epsilon \l(1 - A \r) \mathcal Z
\label{eq:main-eq-rs} 
\end{equation}
with
\begin{equation}
\mathcal Y = R_{\Sigma} - \frac{D_1 D_2}{\sigma^2}
\end{equation}
and
\begin{equation}
\mathcal Z =  2 R_{ab} n^a n^b + K_{ab} K^{ab} - K^2 + \frac{D_1 D_2}{\epsilon \sigma^2} + \frac{2D_{-1}}{\sqrt{\epsilon \sigma^2}} \l( K - \frac{D_1}{\sqrt{\epsilon \sigma^2}} \r)
\end{equation}
where 
\begin{equation}
\sqrt{\epsilon \sigma^2} \; K_{ab} = \sqrt{\epsilon \sigma^2} \; \nabla_a n_b=\nabla_a \nabla_b \l( \sigma^2/2 \r) - \epsilon n_a n_b
\end{equation}
is the extrinsic curvature of the $\sigma^2=$constant foliation, $R_{\Sigma}$ is its induced scalar curvature: 
\begin{equation}
R_{\Sigma} = R - \epsilon \l( 2 R_{ab} n^a n^b + K_{ab}K^{ab} - K^2 \r)
\end{equation}
and we have introduced the convenient notation $D_k=D-k$. Also, $\epsilon = g_{ab} n^a n^b = \pm 1$ 
with $n^a = g^{ab} n_b$. (See Appendix \ref{app:geom-struc} on a more geometrical form of the above expression.)

The  \eq{eq:main-eq-rs} is a non-local expression for the biscalar  $\rmod(p;P,\lp)$. The final step of the analysis is to  extract a local scalar object from this biscalar by taking the coincidence limit. 
In general, to study  any particular curvature invariant $\mathcal{K}(P)$ obtained from the qmetric (using the same algebraic expression that connects the corresponding scalar with the original metric), we start from the modified {\it biscalar} curvature invariant 
${\mathcal{K}}_{(q)}(p;P,\lp)$ evaluated for the qmetric, and take the coincidence limit $\sigma^2 \rightarrow 0$ to obtain the corresponding quantity.
We shall  focus on spacetime regions with {\it regular curvature}. 
Then we can use, for  $\nabla_a \nabla_b \sigma^2$  
the well known expansion  in a covariant Taylor series around the base point $P$ given (see,  e.g., \cite{christenson-worldf}) by: 
\begin{eqnarray}
\frac{1}{2}\nabla_a \nabla_b   \sigma^2  = g_{ab} - \frac{\lambda^2 }{3} \mathcal{S}_{ab} + \frac{\lambda^3}{12} \nabla_{\bm n} \mathcal{S}_{ab} - \frac{\lambda^4}{60} \l(\nabla^2_{\bm n} \mathcal{S}_{ab} + \frac{4}{3} \mathcal{S}_{ia} \mathcal{S}^i_{\phantom{i}b} \r) + O(\lambda^5)
\end{eqnarray}
where $\lambda$ is the arc-length along the geodesic which, of course, is numerically same as $\lambda=\sqrt{\epsilon \sigma^2}$, $\nabla_{\bm n} \equiv n^i \nabla_i$ and $\mathcal S_{ab} = R_{a c b d} n^c n^d$. It is convenient to define a few related quantities as well:
\begin{equation}
\mathcal S = g^{ab} \mathcal S_{ab} = R_{ab} n^a n^b;\quad
\bdot {\mathcal S} = \nabla_{\bm n} \mathcal S; \quad             
\bddot {\mathcal S}= \nabla_{\bm n} \l( \nabla_{\bm n} \mathcal S \r)
\end{equation}
Note that $\mathcal{S}$ is the functional used in the variational principle in the emergent gravity paradigm.  

Carrying out this operation for the Ricci biscalar in $D$ dimensions (see Appendix \ref{app:example} for the details in a special case), we find that we get
an object which depends not only on local tensorial objects such as $g_{ab}, R_{abcd}$ etc at $P$, but also on a vector $n^a$ which, apart from being normalized, becomes arbitrary in the coincidence limit:  
\begin{eqnarray}
\rmod(P,\lp) &=& \underset{p \rightarrow P}{\lim} \rmod(p;P,\lp)
\nn \\
&=& 
\phantom{\frac{\lp^2}{15}} \underbrace{\alpha \l[R_{ab} n^a n^b \r]_P}_{O(1) \; \rm term} 
\;\; - \;\; 
\frac{\lp^2}{15} \;
\underbrace{
\l[ \frac{1}{3} \mathcal S_{ab} \mathcal S^{ab} + \frac{3}{2} \bddot {\mathcal S} + \frac{5}{3} \mathcal S^2 \r]_P
}_{O(\lp^{\;2}) \; \rm term}
\label{eq:mod-rs}
\end{eqnarray}
where $\alpha = 2 \epsilon (D+1)/3$. (In general, there is also a divergent $O(\lambda^{-1})$ term which is well-known in the context of point-splitting regularization. This term can be regularized and dropped; one way to do so formally is described in Appendix \ref{app:div-term}.) This is our final result.

Note that, after the coincidence limit has been taken, the vector $n^a$ on the RHS above must be treated as an arbitrary (normalized) vector. In other words, the object $\rmod(P,\lp)$ at any event $P$ must be treated as a quantity which depends on local tensorial objects at $P$, such as $g_{ab}, R_{abcd}$ etc., as well as on arbitrary normalized  vectors $n^a$. In this sense, we end with an object which depends, in addition to standard geometric objects such as $g_{ab}, R_{abcd}$ etc., on arbitrary vector degrees of freedom at each spacetime event. These vectors have to be thought of as the vestige of the small scale structure of spacetime, and will allow us to connect up with the emergent gravity paradigm; we will come back to this point, in a broader context, shortly.

The most important aspect of the result, as far as this paper is concerned, is the following:
When we carry out
 the last step  (3) of the procedure outlined in   page \pageref{page:ricci},
by taking the $\lp\to0$ limit of \eq{eq:mod-rs}, we  obtain the grin of the Cheshire cat:
\begin{equation}
 \lim_{\lp\to0}\lim_{p\to P}\rmod(p,P;\lp^2)=\alpha R_{ab}(P)n^an^b\propto \mathcal{S}(P) 
\label{lim1}                                                                                                                                                                                                                                                                                                                                                                                                                                                                                  
\end{equation}
In other words,
\begin{eqnarray}
\rmod(P,\lp=0) \neq R(P)
\end{eqnarray}
That is, we start with Ricci biscalar for the qmetric (which bears the same algebraic relation to qmetric as the usual Ricci scalar does to the usual metric), take the coincidence limit $p\to P$ and then take the `classical limit' of $L_P\to0$; only to find that the resulting expression is not\footnote{As a curious aside we mention the following: The condition $\alpha R_{ab}n^an^b = R$  numerically, requires the rather peculiar relation $R_{ab} = 3/[2(D+1)] g_{ab} R$ to be satisfied by the background metric. Even the maximally symmetric space(time)s satisfy this relation only for 
the special case $D=2$; this includes,  e.g., the $2$-sphere.}  the Ricci scalar for the background metric!
		
On the other hand, it is easy to see from Eq. (\ref{eq:main-eq-rs}) that if we  take the `classical limit' \textit{first}, we get:
\begin{eqnarray}
\rmod(p;P,\lp=0) = R(P)
\end{eqnarray}
Clearly, the limits $\sigma^2 \rightarrow 0$ and $\lp \rightarrow 0$ do not commute: 
\begin{equation}
	\lim \limits_{\lp\rightarrow0} \lim \limits_{\sigma^2\rightarrow0} \rmod(p;P,\lp) \; \neq \; \lim \limits_{\sigma^2\rightarrow0} \lim \limits_{\lp\rightarrow0} \rmod(p;P,\lp)
\label{nc1}
\end{equation}
The origin of this non-commutativity of the limits can be traced back to the factor $\l(1 - A^{-1}\r)$ in the second term of Eq. (\ref{eq:main-eq-rs}); in fact, the function
		\begin{eqnarray}
		f(\sigma,\lp) = 1 - A^{-1} 
		= 1 - \frac{1}{1 + \lp^2/\sigma^2}
		\end{eqnarray}
		has \textit{no limit at $(\sigma,\lp)=(0,0)$}, since
		\begin{equation}
		\underbrace{\lim \limits_{\lp\rightarrow0} \lim \limits_{\sigma^2\rightarrow0} f(\sigma,\lp)}_{=1} 
		\; \neq \; 
		\underbrace{\lim \limits_{\sigma^2\rightarrow0} \lim \limits_{\lp\rightarrow0} f(\sigma,\lp)}_{=0}
		\end{equation}
		In the spirit of conventional regularization techniques (in our case, the most natural one is, of course,  {\it point-splitting} regularization), we must take the coincidence limit $p\to P$  first, which results 
		in the structure of $\rmod(P)$ given by \eq{eq:mod-rs}.
	{\it The resulting expression in \eq{lim1}, arising as the leading order term is  our key result.} 

Given the rather surprising and counter-intuitive nature of the result, we give an explicit demonstration of the same, in a simpler context, Appendix \ref{app:example}. 
This is done by using the synchronous coordinate system for the background metric and taking $p$ and $P$ along radial direction in the Euclidean space. This is, of course, not the most general case but this captures much of the subtleties in the calculation. The interested reader is referred to 
Appendix \ref{app:example} 
for the details as to how the non-commuting nature of the limits lead to this result. 

We shall now discuss the implications of our result  for the emergent gravity paradigm.

\section{Implications} \label{sec:implications}
\subsection{The  emergent gravity variational principle} \label{sec:impl} 
The standard action for general relativity is based on the Einstein-Hilbert Lagrangian: $\mathcal L_{\rm EH} = R$. The usual belief is that one has to somehow quantise the theory based on this Lagrangian. But if gravity is an emergent phenomenon like elasticity, this effort is like quantizing a theory based on a Lagrangian describing elastic vibrations. Obviously, this will not  lead to lasting progress. 

The problem with the emergent gravity program, on the other hand, is that it a `top-down' approach (in length scales) and the next logical step of the programme is something like trying to discover statistical mechanics from thermodynamics. \textit{One simply cannot do this without additional postulates.} Further, what we are really interested is in  the quasi-classical limit in which ideas like differential manifold, metric etc remain valid and only the dynamical description changes. In this limit,  the modification in \eq{funda} captures some of the key aspects of quantum gravity. This modification, in turn, is equivalent to using the qmetric in place of the original metric. 

Given such a replacement, the most natural extremum principle will be the one based on the Ricci biscalar $\mathcal{R}(p,P; L_P^2)$ for the qmetric, just as we think of the extremum of the usual Ricci scalar $R$ (for the usual metric) as the natural choice in general relativity. To obtain a local variational principle from this biscalar we take the limit of $p\to P$ in $\mathcal{R}(p,P; L_P^2)$, obtaining the result in \eq{eq:mod-rs}. That is, in the limit of $p\to P$, the extremum principle based on $\mathcal{R}$ is equivalent to the extremum principle based on the limiting expression (viz. the first term in \eq{eq:mod-rs} when $L_P 
\to 0$), which is just $\mathcal{S} \propto R_{ab}(P)n^an^b$. The variation of the metric, which is equivalent to the variation of geodesic distance, translates into varying $n^a$ in this expression since we treat the base point $P$ as fixed.
In other words, there is a natural transmutation of the variational principle when we carry out the steps 1-3 outlined in page \pageref{page:ricci}. We see that:
\begin{equation}
\mathcal L_{\rm EH} = R \;\;\; \longrightarrow \;\;\; \mathcal L_{\rm eff} = \alpha R_{ab} n^a n^b
\label{tr} 
\end{equation}
where $\bm n$ is a vector of constant norm. 
The $\mathcal L_{\rm eff}$ has precisely the form of the  {\it thermodynamic functional} $\mathcal{S}$ first suggested by Padmanabhan et al. \cite{entropy-functional} as a basis for an alternate variational principle for describing emergent gravitational dynamics, motivated by thermodynamics of causal horizons \cite{dk-ig}. One can treat this as a functional of the normalised vector field $n^a$ and the extremum must now hold for variations of all $n^a$. As demonstrated in several previous papers \cite{grtp,entropy-functional},  this leads to standard field equations of gravity with the \cc\ arising as an undetermined integration constant.\footnote{The idea of gravity being described fundamentally by a {\it non-local object} with the {\it geodesic distance} playing the key role (which was recently emphasized in \cite{dk-ml}) 
seems to be conceptually in tune with some earlier work by Alvarez et al. \cite{alvarez-etal}. In our framework, the geodesic distance appears naturally if spacetime has a built in zero-point length which does not violate Lorentz invariance.}

The results in \eq{eq:mod-rs} and \eq{lim1} are  highly non-trivial and we could not have guessed them from any simple consideration.
These results highlight the {\it robustness} of the framework for emergent gravity paradigm based on an alternate variational principle by demonstrating that such a principle could be {\it enforced} upon us by the existence of a fundamental length scale in spacetime. The conceptual loop closes nicely since the emergent gravity paradigm was itself largely motivated by existence of thermal attributes of causal horizons. One of these attributes, the horizon entropy, however, turns out to be {\it divergent} when viewed as entanglement entropy due to tracing over the vacuum fluctuations of fields beyond a causal horizon. It was argued in \cite{tp-pid-cutoff} that this divergence can be removed by the very same introduction of a minimal length scale which renders the two-point correlators of the quantum fields finite. Our result shows that incorporating such a minimal length scale in turn leads to an extremum principle which incorporates the thermodynamic features of gravity in a natural manner.

\subsection{Implications for the cosmological constant problem}

One major conceptual bonus we obtain from the emergent gravity paradigm is the possibility of understanding the cosmological constant. It has been argued \cite{grtp,hptp}  that a clean solution to \cc\ problem can be obtained \textit{only if} the metric is not treated as a dynamical variable in a local extremum principle.  The thermodynamic variational principle motivated by emergent gravity paradigm satisfies this condition and obtains the equations by varying a vector field of constant norm. In this approach, the \cc\ arises as an undetermined integration constant in the solution \cite{grtp,hptp}. 

Here, we have shown that such a variational principle can indeed arise as a non-trivial limit of a calculation that incorporates the zero-point-length.
The emergent gravity paradigm also suggests a particular approach towards determining the numerical value of the \cc\ \cite{hptp}. This, in turn, depends on the existence of a minimal area $L_P^2$ in quantum gravity so that the number of surface degrees of freedom of a sphere with radius $L_P$ is given by $N_{\rm sur} = (4\pi L_P^2/L_P^2) = 4 \pi$. The existence of a zero-point length, on which entire analysis of this paper is based, ties in nicely with the existence of a minimal area and a minimal count for the surface degrees of freedom. This gives the hope that a more sophisticated model will be able to put these results on a firmer footing.

\section{Discussion and Outlook} \label{sec:outlook}

\begin{table}
\begin{center}
\begin{tabular}{p{0.3\textwidth}p{0.3\textwidth}p{0.35\textwidth}}
\hline\hline
\noalign{\smallskip}
Let $L_P^2 \to 0$ with $p\neq P$ (no surprises here!) & The Strategy of this paper & Let $p\to P$ with $L_P^2 \neq 0$ (leads to entropy density of emergent gravity paradigm)\\
\noalign{\smallskip}
\hline\hline
\noalign{\medskip}
$\sigma^2 \to \sigma^2 (P,p)$ & (1) Start with the geodesic interval $\sigma^2 (P,p)$ for a metric $g_{ab}$ & $\sigma^2 \to 0$\\
\noalign{\medskip}
$\sigma_{(q)}^2 \to \sigma^2 (P,p)$ & (2) Incorporate some QG effects by the ansatz $\sigma_{(q)}^2 (P,p, L_P^2) = \sigma^2 (P,p) + L_P^2$ & $\sigma_{(q)}^2 \to L_P^2$\\
\noalign{\medskip}
$q_{ab}(P,p, L_P^2) \to g_{ab} (P)$ & (3) Find the qmetric $q_{ab}(P,p, L_P^2)$ related to $\sigma_{(q)}^2 (P,p, L_P^2)$ & Diverges as $(L_P^2/\sigma^2)|_{\sigma \to 0} $ \\
\noalign{\medskip}
$R(P,p, L_P^2) \to R(P)$ & (4) Compute the Ricci biscalar $R(P,p, L_P^2)$ for the $q_{ab}(P,p, L_P^2)$ & \fbox{$R(P,p, L_P^2) \to \mathcal{S} + O (L_P^2)$} \\
\noalign{\medskip}
\hline
\end{tabular}
\caption{Summary of the paper. The strategy adopted in the paper is described in the middle column with the logical flow being from the top to the bottom. We take the coincidence limit to obtain local quantities from bitensors, biscalars etc. These limits are shown in the  right column. As a crosscheck, we have included a left column showing what happens when we set $L_P^2=0$. (The results in this column are trivial and as expected.) See text for more discussion.}
\label{table:strategy}
\end{center}
\end{table}

The Table \ref{table:strategy} summarizes the results of the paper in a thematic manner. The analysis involves four key steps, which are given in the \textit{middle} column as steps 1-4 and described in detail in the paper. 

\textit{Step 1}: We choose to describe a classical spacetime, not in terms of a metric, but in terms of its geodesic interval, which is a biscalar and contains exactly the same information as the metric. Upgrading the role of $\sigma^2(P,p)$ as \textit{the} descriptor of geometry is \textit{a key new aspect} of this paper.

\textit{Step 2}: We recall that certain aspects of quantum gravity can be incorporated by the ansatz $\sigma_{(q)}^2 (P,p, L_P^2) = \sigma^2 (P,p) + L_P^2$. This description is valid in the quasi-classical region between the full quantum gravity domain (in which we do not know what replaces a differential manifold, metric etc) and the classical domain (in which conventional GR holds). We expect the quasi-classical region to admit a description in terms of  effective geometrical variables built from the qmetric.

\textit{Step 3}: We next determine the form of a symmetric, second rank, bitensor $q_{ab}(P,p, L_P^2)$ which corresponds to the modified geodesic interval $\sigma_{(q)}^2 (P,p, L_P^2)$. (See Sec. \ref{sec:rs}). It bears the same relation to $\sigma_{(q)}^2$ as the background metric $g_{ab}$ bears to $\sigma^2$ and serves as the analogue of the metric in the quasi-classical domain. 

\textit{Step 4}: We compute the Ricci biscalar $R(P,p, L_P^2)$ for the qmetric $q_{ab}(P,p, L_P^2)$. The local object $R(P, L_P^2)$ obtained by $p\to P$ limit on this biscalar is a good candidate for the variational principle in quasi-classical domain. We find that the leading order, $L_P$-independent term in $R(P, L_P^2)$ is \textit{given by the functional $\mathcal{S}$ used previously in the thermodynamic variational principle in the emergent gravity paradigm!}  This is our key result.

In the right column, we have given the coincidence limit $p\to P$ of various quantities keeping $L_P^2$ nonzero. To begin with, we see that, in the coincidence limit $\sigma^2\to0$ while 
$\sigma_{(q)}^2\to L_P^2$, \textit{by design} of our ansatz. It is this difference which is at the foundation of our result. The coincidence limit of $q_{ab}$ is divergent, which is easy to understand because we do not expect notions like metric  to survive when $\sigma^2\lesssim L_P^2$; this is encoded in the factor $(L_P^2/\sigma^2)$ in $q_{ab}$. Finally, when we take the coincidence limit of the 
Ricci biscalar $R(P,p, L_P^2)$ for the qmetric, we get the final result shown in the boxed equation at bottom right.
Just as a crosscheck, we have given the $L_P^2\to0$ limits on the left column which are trivial and behave as expected.

Our analysis raises several important points which need further investigation, and their clarification will provide further insights into the effects of minimal length on small scale structure of spacetime. We will now describe some of them.

\textit{1. Can one trust the grins of the Cheshire cats?}
 
The key step which leads to this result is the possibility that effects of a minimal length scale are not necessarily ``small" and vanish when $L_P\to0$ limit is taken. While this might appear rather surprising, we do know of such results in other areas of physics. 
We shall briefly describe a few of such examples to present this concept in a broader context.

The simplest example of such a result arises in the study of electrons in, say, helium atom. If one solves the exact non-relativistic Schr\"odinger equation for the two electrons in a helium atom, one will, in general, obtain a wave function $\psi (\mathbf{x}_1, \mathbf{x}_2)$. We will, however, find that among all such functions which satisfy the Schr\"odinger equation, only half of them --- viz., those which satisfy the antisymmetric condition $\psi (\mathbf{x}_1, \mathbf{x}_2) = -\psi (\mathbf{x}_2, \mathbf{x}_1)$ --- occur in nature. No amount of study of the Schr\"odinger equation for helium atom will explain the peculiar phenomena related to the Fermi statistics. The actual explanation lies deeply buried in the relativistic quantum field theory from which we can obtain the Schr\"odinger equation by a suitable limiting process involving the $c\to \infty$ limit. In this limit, $c$ disappears from the relevant equations but a peculiar feature of the relativistic quantum field theory remains as a leading order (i.e., $c=0$ limit) relic in the non-relativistic helium atom. What is more, this effect is quite different from, say, the usual relativistic ``corrections'' to Schr\"odinger equation which will admit a Taylor series expansion in $(1/c)$; we cannot say that Fermi statistics is obeyed with increasing accuracy in a similar Taylor series expansion in $(1/c)$! It is a completely different kind of low energy relic from the high energy theory. 

A moment of thought will show that this is very similar to what happens in our case. When $L_P \to 0$ limit is taken in a result incorporating certain quantum gravitational effects, we obtain a residual relic viz., the transmutation of the variational principle indicated in \eq{tr} which survives with no trace of Planck length. Moreover, we cannot think of this as a Taylor series in $L_P^2$ because the leading order term itself is different. But, this should not be conceptually any more surprising than the behaviour of wave functions of helium atoms. In neither case can one guess, staring at the low energy theory, the origin of the relic. 

A slightly more technical result of this kind was discussed in Ref.~\cite{tphpqftnr}. It was shown that, when one proceeds from the action functional for  a relativistic particle to the action functional for the non-relativistic particle, in the path integral approach to quantum theory, a relic term arises in the $c\to \infty$ limit leading to a well-defined phase factor in the non-relativistic wave function. Once again, the origin of this phase factor (which is independent of $c$ though arises as a relativistic relic) in the non-relativistic limit is completely mysterious if one did not know the relativistic version of the relevant expressions.

Mathematically, our result arises from the non-commutativity of the limits in \eq{nc1}. Such effects are also known in the literature and we give two examples. First arises in computing curvature tensors of some rather simple metrics. Consider a metric $g_{ab}(x;\varepsilon)$ which depends on some parameter $\varepsilon$, from which we compute the curvature tensor $R_{abcd}(x;\varepsilon)$ and then take the limit $\varepsilon\to0$. We now compare this result with the one obtained by first taking the limit $\varepsilon \to0$ in the metric and then computing the curvature. While these two procedures will usually give the same result, it is possible to construct \cite{tptopo} several simple metrics in which they do \textit{not}, with the difference being a singular term (usually a Dirac delta function). 

A more technical result of similar nature arises due to interplay between loop divergences in conventional quantum field theory, and attempts to regularize them using regulators which break Lorentz symmetry, as was first highlighted in Refs.~\cite{collins-etal-lv}. These authors pointed out the inevitability of $\mathcal{O}(1)$ effects arising due to Lorentz violations (LV) at higher energies, whose effects can generically get dragged  to lower energies unsuppressed, due to interplay between {\it radiative corrections}, which involve large loop momenta $k$, and the fact that the LV terms are also expected to {\it regulate} the UV divergences. In a sense, this result is also  a consequence of the non-commutativity of the two limits, corresponding to whether one sets the regulating function to zero before or after evaluating the loop integral. Our result points to something similar in the context of small scale ``geometry" of spacetime, where we have retained Lorentz invariance but have abandoned {\it locality}. (The key difference,  of course, is that in our case, the $\mathcal{O}(1)$ term we obtain turns out to have important physical implications.)

Finally, even the familiar conformal anomaly, arising in theories which are classically conformal invariant --- and, in fact, many other symmetry breaking anomalies --- can be thought of as quantum residues of  similar nature. If we regularize a theory by dimensional regularization, we use expressions in $D$ dimensions  and finally take, say, the $D\to4$ limit. The theory which is conformally invariant in $D=4$ will not, in general, be conformally invariant in  
$D\neq 4$ and when we eventually take the limit $D\to4$, we get an anomalous result. This result, again, is difficult to understand working entirely in the $D=4$ situation but is clear when we think of it as a limiting process leaving a residue.

These examples show that  when certain limits are taken in a theoretical model the resulting theory could contain relics of the more exact description. In all such cases \textit{no amount of study of the approximate theory will give us a clue as to where the relic came from} (e.g., the study of Schr\"odinger equation for the helium atom can never lead us to the Fermi statistics for the electrons). We believe the transmutation of the variational principle in \eq{tr} is of similar nature which is nearly impossible to understand within the context of classical gravity itself. Our analysis throws light on this and shows that it could be a valid relic of quantum gravity.

\textit{2. How valid is the assumption that quantum gravity effects would modify length scales as: $\sigma^2 \rightarrow \sigma^2 + \lp^2$? }
	
This was the key input for deriving the qmetric in \cite{dk-ml} and forms the starting point of the entire analysis.  There has been a good deal of evidence in favor of this modification, from several independent lines of analyses dating back to $1960$'s. The two aspects which need scrutiny in this regard are the following: (a) Can a c-number description with the replacement
$\sigma^2 \rightarrow \sigma^2 + \lp^2$ capture some of the quantum gravitational effects?
(b) How unique is such a replacement to bring in \eq{zpl1} which probably is a more precisely stated result?

The first issue is similar to what is encountered in other contexts as well. As a simple example, consider a  harmonic oscillator with the Hamiltonian $H(p,q)=(1/2)[p^2+ q^2]$ (in convenient units). Classically $H$ has minimum at $p=0=q$ corresponding to the ground state in which the oscillator sits at the minimum of the potential with zero velocity. Quantum mechanically, uncertainty principle prevents us from giving precise values to $q$ and $p$ simultaneously and hence this cannot be the description of the ground state. The exact analysis of this problem, of course, will involve treating $H$ as a operator and finding its lowest eigenvalue etc. However we can take the point of view that when $q$ is close to zero, the uncertainty in the momentum is $\hbar/q$ and the relevant c-number to minimize is a `quantum-corrected' Hamiltonian $H_{qc}\equiv (1/2)[(\hbar/q)^2+q^2]$ with the minimum value being $\mathcal{O}(\hbar)$.
We can indeed capture the essential feature by just making the replacement $H\to H_{qc}$. We consider the replacement $\sigma^2 \rightarrow \sigma^2 + \lp^2$   to be similar in spirit, getting us some quantum results at c-number price! Note that the existence of the zero-point-length can be demonstrated by considering the effect of the uncertainty principle  on spacetime measurements \cite{zpltp}, so the analogy above is conceptually quite close. 

As regards question (b), one could say the following:  The ansatz $\sigma^2 \rightarrow \sigma^2 + \lp^2$, of course, is not the only modification possible to incorporate \eq{zpl1}, 
and in fact we show in Appendix \ref{app:generalized-qmetric} how to get the corresponding results for a more general modification $\sigma^2\to S(\sigma^2)$. But in general, the resulting form of the qmetric will not reduce to flat spacetime if the original spacetime is flat. The specific choice $\sigma^2 \rightarrow \sigma^2 + \lp^2$
has the additional advantage that it reduces the qmetric to flat spacetime when the original metric is flat. In this sense, the modification we have worked with is indeed special (and might even be unique). Consider, e.g., another modification \cite{brown-qg}
: $\sigma^2 \rightarrow \sigma^2 {\mathcal F}
(\sigma^2/\lp^2)$ where ${\mathcal F}(x) = e^{1/x}$ (a naive expansion in $\lp^2$ gives $\sigma^2 \sim \sigma^2 + \lp^2$). This is another example of a deformation which is non-analytic at $x=0$.  It 
would be interesting to see if there exists a specific sub-class of such functions which leaves a $\mathcal{O}(1)$ effect on curvature invariants; work along these lines is in progress. 

\textit{3. Does analytic continuation from the Euclidean to Lorentzian spacetime lead to special difficulties?}

Not really, in the mathematical sense. In the Euclidean sector a constant norm vector is always `spacelike' (we use the signature $(-, +, +, +, \ldots)$ in the Lorentzian spacetime) while in the Lorentzian spacetime it can be timelike, spacelike or null. The previous results in emergent gravity paradigm are usually presented for null vectors but the results continue to hold for any constant norm vector. The introduction of the zero-point-length and its manipulations are best done, however, in the Euclidean space which is what we have done. It is easy to see that the results hold without any ambiguity for both spacelike and timelike vectors in the Lorentzian space. Therefore one can take the null limit by a continuity argument after straddling the null surface by spacelike and timelike vectors on the two sides with some care  in working with the affine 
distance along a null ray.
	
\textit{4. To what extent can we think of the qmetric $\gm_{ab}(p,P)$ as  a metric?  }

This issue is somewhat  irrelevant to our results and --- more generally --- when one treats a {\it distance function} $d(x,y)=\sqrt{\epsilon \sigma^2}$ defined on a manifold as more fundamental than the metric, and the metric $g_{ab}(x)$ as a derived quantity  
	which enables us to construct geometric invariants for a manifold. 	
	As explained in detail in \cite{dk-ml}, the non-local character of the qmetric is simply a physical characterization of quantum fluctuations which 
	leave their imprint (in this approach) in the form of a lower bound on the intervals: $d(x,y) \geq \lp$. 
	At the smallest of the scales, we do not expect any local tensorial object to describe the quantum spacetime geometry accurately anyway, and hence it is not 
	unexpected that more general mathematical objects must replace the conventional ones. \textit{The ideas here can be thought of as a first step in this direction.}
	In fact, non-local effective actions have been considered for quite some time in the context of quantum gravity (e.g., DeWitt proposed such an action in \cite{universal-reg}), and $\gm_{ab}(p,P)$ might serve as 
	an important mathematical object to build such actions. 
	
\textit{5. What are the implications for spacetime singularities?}
	 
	Some preliminary results along these lines were given in \cite{dk-ml}. But to answer the question in full generality, one needs to obtain the full curvature tensor for the qmetric and then study typical curvature invariants. (We expect to do this 
	in a future work.) Even restricting to the Ricci scalar (which we do have), one needs to revert to the general equation, Eq.~(\ref{eq:main-eq-rs}), instead of Eq.~(\ref{eq:mod-rs}) which is arrived at by assuming that one is 
	working in regions of finite curvature. The transition from Eq.~(\ref{eq:main-eq-rs}) to Eq.~(\ref{eq:mod-rs}) makes use of the covariant Taylor series expansions of various bitensors involved, and the coefficients of such 
	series depend on the curvature and its derivatives. If the latter blow up, as they are expected to near spacetime singularities, then one must be careful about using such series. Efforts are ongoing to evaluate 
	Eq.~(\ref{eq:main-eq-rs}) for some physically relevant singular spacetimes using valid expressions for the world function.

\textit{6. What about the $\mathcal{O}(\lp^2)$ term ?}

Since the leading order term in \eq{eq:mod-rs} gives the entropy functional of emergent gravity, (leading to classical Einstein's equations) one might be tempted to consider the next term as some kind of quantum gravitational correction and explore its consequences. This temptation must be resisted for the following reason: The entire philosophy behind the analysis was that classical spacetime is a non-perturbative limit of quantum spacetime and it is incorrect to do a perturbative model of quantum gravity. So, when the leading order ($L_P^2=0$) term itself is not the classical term, it is inconsistent to study quantum corrections perturbatively, order by order in $L_P^2$. The second, rather technical reason, for not pursuing this line of attack has to do with the fact that we are so far dealing with pure gravity. We do not know how to get the matter stress tensor from a corresponding quantum calculation by taking a suitable limit. (In emergent gravity one can simply add the thermodynamic potential of matter, $T_{ab}n^an^b$  and get the correct result but we do not have, yet, the corresponding quantum version.) So it is rather useless to compute quantum gravitational corrections in source-free spacetimes. (If one does this, in spite of these reservations, one finds that $R_{ab}=0$ continue to be a solution even with the lowest order quantum correction terms.)  

The key insight coming from our analysis, going beyond the specifics,  is the following: It seems likely that non-local but (hopefully) Lorentz in(co)variant deformation of the spacetime geometry at small scales, will be  a consequence of quantum fluctuations of the (as yet unknown) microscopic degrees of freedom of quantum gravity. This leads to an
 inevitable  $\mathcal{O}(1)$ modification due to a minimal spacetime length.  Regardless of the precise form of the deformation, the results will now essentially depend on the two limits, $\xi \rightarrow 0$ and $\xi \rightarrow \infty$ (with $\xi=\sigma^2/\lp^2$) being inequivalent. Any dimensionless deformation function $\mA$ can only depend on $\xi$; 
$\mA=A[\xi]$. Hence, unless $A[\xi]$ has same limit at $\xi \rightarrow 0$ and $\xi \rightarrow \infty$, (e.g., it is symmetric under $\xi \rightarrow \xi^{-1}$) one will generically obtain the kind of non-trivial result that we have obtained. The precise form of the  $\mathcal{O}(1)$ term would depend on the form of $A[\xi]$; our choice, $A[\xi] = 1+ 1/\xi$, yields a residual term which happens to connect up with certain key ideas concerning emergent gravity and cosmological constant that have been developed over the past decade.

\section*{Acknowledgments}
 Various expression(s) given/used in this work have been verified for special cases by symbolic computations in \texttt{Maple} using GRTensorII \cite{grtensor}. The research work of TP is partially supported by the J.C.Bose Fellowship of DST, India. DK thanks IUCAA, Pune, where part of this work was done, for hospitality. We thank the referee for several constructive comments. 
\appendix 
\section{Determination of qmetric associated with arbitrary modifications of geodesic
intervals} \label{app:generalized-qmetric}
It is possible to generalize the qmetric approach to the cases in which the geodesic
intervals are modified in an arbitrary manner. We briefly comment on this
possibility in this appendix, leaving a
detailed analysis of such a generalization for future work.

Let us consider the case in which the geodesic intervals are modified as $\sigma^2 \to
S(\sigma^2)$ and the scalar propagator is correspondingly modified as $(\sigma^2)^{-(D-2)/2}\to
(S(\sigma^2))^{-(D-2)/2}$ where
 $S(\sigma^2)$ is a given function.
(The choice $S(x)=x+\lp^2$ will reproduce the results of this paper). We want to determine the form of a qmetric $\gm^{ab}(p,P)$, (which is a second rank, symmetric, bitensor) built out of $\sigma^2(p,P), g_{ab}$ and $n^a=g^{ab} n_b$, such that the following two conditions are satisfied:
(i) The modified geodesic interval $S$ satisfies \eq{HJ1} with the metric replaced by the qmetric and (ii) The modified propagator $(S(\sigma^2))^{-(D-2)/2}$ will satisfy the Green function equation with the metric replaced by the qmetric. This is indeed possible and we outline the steps here:

Let us
assume that such a qmetric has the form
\begin{eqnarray}
\gm^{ab} = A^{-1} g^{ab} + \epsilon Q n^a n^b=A^{-1} h^{ab} + \epsilon (A^{-1}+Q) n^a n^b
\end{eqnarray}
where $A$ and $Q$ are (as yet arbitrary) functions of $\sigma^2$ and $h^{ab} =  g^{ab} - \epsilon n^a n^b$ is the induced metric on the
hypersurface with normals $n^a$. This ansatz is motivated by the fact that qmetric $\gm^{ab}$ is a second rank, symmetric, bitensor built from $g^{ab}$ and $n^a$. 
Note that the corresponding covariant components are given by:
\begin{equation}
 q_{ab}=Ag_{ab}-\epsilon B n_a n_b;\qquad B=QA^2(1+QA)^{-1}
\end{equation} 
We want
$S$ to be the geodesic interval for $\gm^{ab}$, and hence we substitute this {\it
ansatz} in the Hamilton-Jacobi equation:
\begin{eqnarray}
\gm^{ab} \partial_a S \; \partial_b S = 4 S
\label{condHJ}
\end{eqnarray}
This gives, on using $g^{ab} \partial_a \sigma^2 \; \partial_b \sigma^2
= 4 \sigma^2$, the following relation between coefficients $A, Q$ in the
metric and the function $S$:
\begin{eqnarray}
A^{-1} + Q = \frac{1}{\sigma^2} \frac{S}{S'^2}
\end{eqnarray}
where $S'=\DM S/\DM \sigma^2$. Obviously, the condition in \eq{condHJ} can only fix the projection of the qmetric
in the subspace spanned by $n^an^b$; so we have:
\begin{eqnarray}
\gm^{ab} = A^{-1} h^{ab} + \epsilon \l( \frac{1}{\sigma^2} \frac{S}{S'^2} \r) n^a n^b
\end{eqnarray}
 To fix the form of $A$ we use the condition on the Green function. For an arbitrary function $S(\sigma^2)$,
it is readily shown that the flat space propagator $G(\sigma^2)$ will be modified to $G(S(\sigma^2))$ provided that $A=S(\sigma^2)/\sigma^2$. 
To establish this, one uses the above form of the qmetric and the relation
\begin{eqnarray}
\sqrt{\rm |det \; \gm |} = \frac{ A^\frac{D-1}{2}}{\sqrt{A^{-1} + Q}} \sqrt{|g|}
\end{eqnarray}
which follows from the {\it matrix determinant lemma}: 
\begin{equation}
{\mathrm{det}} \l(\rm \bf M + \rm \bf u \rm \bf v^{\mathrm{T}}\r) = \l(\mathrm{det} \; \rm \bf M\r) \times \l( 1 + \rm \bf v^{\mathrm{T}} \rm \bf M^{-1} \rm \bf u \r)                                                                                                                                                                         \end{equation} 
where $\rm \bf M$ is an invertible square matrix, and $\rm \bf u, \rm \bf v$ are column vectors (of same dimension as $\rm \bf M$).
Subsequent computation of $\square_{(\gm)} G(S(\sigma^2))$ is straightforward (although lengthy). Enforcing the condition that $\square_{(\gm)}  G(S(\sigma^2))=0$ for $p\neq P$ gives the differential equation
\begin{eqnarray}
\frac{\DM \ln A}{\DM \ln \sigma^2} = \frac{\sigma^2 S'}{S} - 1
\end{eqnarray}
which has the solution: $A=S(\sigma^2)/\sigma^2$. (The multiplicative constant is fixed by the condition that $A=1$ when $S(\sigma^2)=\sigma^2$). 

This fixes the final form of the qmetric to be:
\begin{eqnarray}
\gm^{ab} &=& \frac{\sigma^2}{S(\sigma^2)} h^{ab} + \epsilon \l(\frac{1}{\sigma^2} \frac{S}{S'^2}  \r) n^a n^b 
\end{eqnarray}
For the choice made in this paper, $S=\sigma^2 + \lp^2$, one recovers \eq{eq:key1}. For an
arbitrary $S(\sigma^2)$, the resultant qmetric can (generically) yield a
non-zero curvature even when $g_{ab}$ represents a flat spacetime, which does \textit{not} happen for the choice $S=\sigma^2 + \lp^2$. The above formulation is
quite promising in yielding a considerable generalization of the results
in \cite{dk-ml}, and is currently being pursued.

\section{Evaluation of Ricci scalar for the qmetric} \label{app:rs-eval-details}

One can employ a nice trick based on the Gauss-Codazzi decomposition to obtain the Ricci scalar for the modified metric. In fact, the same trick works also for the conformally related metrics, and we outline this case first 
since the logic remains the same. 

Suppose two metrics are related by a conformal transformation: $\tilde g_{ab}=F(x) g_{ab}$. Then, one can imagine foliating the spacetime by vector fields normal to $F=$ constant surfaces, given by $n_i=\nabla_i F(x)/\sqrt{ \epsilon \nabla_a F \nabla^a F }$. The conformal transform of this vector field will  be $\tilde n_i = \sqrt{F} n_i$, whereas $\tilde n^i = n^i/\sqrt{F}$ where $n^i=g^{ij} n_j$. (Note that one needs to be careful about indices; all indices on $\bm{\tilde n}$ are raised or lowered using $\tilde g_{ab}$.) 

The advantage of doing this is that the induced geometries of this foliation in the two metrics are related in a simple manner. The relation between the extrinsic curvatures $\tilde K_{ab}$ and $K_{ab}$ is easy to establish, and turns out to be
\begin{eqnarray}
\tilde K_{ab} = \sqrt{F} K_{ab} + \l( n^k \nabla_k \sqrt{F} \r) h_{ab}
\end{eqnarray}
where $h_{ab}$ is the induced metric. It is easy to show that $\tilde h_{ab}=F h_{ab}$. We now use the Gauss-Codazzi relation
\begin{eqnarray}
R_{(q)} = R_{\Sigma,{q}} - \epsilon \l( \tilde K^2 + \tilde K_{ab}^2 + 2 \tilde n^i \tilde \nabla_i \tilde K \r) + 2 \epsilon \tilde \nabla_i \tilde a^i
\end{eqnarray}
where $\tilde a^i = \tilde n^k \tilde \nabla_k \tilde n^i$. Again, it can shown by straightforward computation that $\tilde a_i=a_i$ and $\tilde a^i=a^i/F$. Most importantly, since we are considering $F=$ constant foliation, 
the induced Ricci scalars are related by a simple scaling: $R_{\Sigma,{q}} = R_{\Sigma}/F$. Putting all these together, it can be shown that one recovers the well known relation between the Ricci scalars of conformally related metrics. 

The above procedure can be generalized for the qmetric. This case is more complicated due to the presence of the $n_a n_b$ term in the metric, but otherwise the steps remain the same. One first shows that, even for this general case, the induced metrics are related by a conformal transformation as above, and the extrinsic curvature $\tilde K_{ab}$ can be written easily in terms of $K_{ab}$ and $h_{ab}$. (These relations are quoted in \cite{dk-ml}). Therefore, one again has $R_{\Sigma,{q}} = R_{\Sigma}/F$. However, the extrinsic curvature terms in the Gauss-Codazzi relation produce a more complicated relation, although there are no new conceptual points involved. The algebraic details, however, are a bit longer, and will be presented in \cite{WIP}.

\section{Explicit demonstration of our result in a special case} \label{app:example}
In this Appendix, we will provide an explicit demonstration of how our key result arises in a special case which captures all the key  features of the general case. We consider a Euclidean spacetime described in the synchronous coordinates with the line element:
\begin{equation}
 ds^2=dt^2+h_{\mu\nu}(t,x^\alpha)dx^\mu dx^\nu
\end{equation} 
This spacetime possesses a geodesic interval function $\sigma(x,x')$ from which \textit{all} the metric coefficients can be obtained using \eq{gR}. It is, however, well-known (see, e.g. \cite{LL2}, page 288) that $x^\mu=$ constant are geodesics in this spacetime and the geodesic interval between 
two points in such a geodesic is just $(t-t')$. So, if we confine our attention to two points $p$ and $P$ along the `radial' direction [for which $x^\mu=$ constant], then we can take the $t$ coordinate as numerically equal to the geodesic interval and write the metric as:
\begin{equation}
 ds^2=d\sigma^2+h_{\mu\nu}(\sigma,x^\alpha)dx^\mu dx^\nu
\end{equation}
Note that we have now downgraded the \textit{function} $\sigma(x,x')$ to a coordinate label and this will work only for the radial geodesics. Our aim is to confine ourselves to points 
$p$ and $P$ along the $\sigma$ direction and illustrate our results when $\sigma\to0$. (This is the difference between the general case and the case we have taken up here for the illustration.)

The qmetric in this case is given by \eq{eq:key1} with $n_adx^a=d\sigma$ and the resulting line element is:
\begin{equation}
ds^2_{(q)}=\frac{d\sigma^2}{A}+Ah_{\mu\nu}(\sigma,x^\alpha)dx^\mu dx^\nu; \qquad A=\l(1+\frac{L_P^2}{\sigma^2}\r)
\label{qmsync}
\end{equation} 
Our task is now straightforward: Compute $R$ for this metric, expand everything related to the background in a Taylor series in $\sigma$ assuming the background spacetime is nonsingular, take the limit of $\sigma\to0$ and identify the leading order term (which we expect to be proportional to $R_{ab}n^an^b$), identify the divergent term (which we expect to be proportional to $1/\sigma$)
and identify the $\mathcal{O}(L_P^2)$ term thereby demonstrating our result. We will now outline the key algebraic steps in this calculation.

Computing $R_{(q)}$ for the line element in \eq{qmsync} is straightforward using the formulas in, say, ref.\cite{LL2} (after making the sign switch $h_{\mu\nu}\to -h_{\mu\nu}$ to go from Lorentzian to Euclidean signature). This gives:
\begin{equation}
R_{(q)} = A^{-1}R + (1- A^{-1}) (R - R_{\Sigma}) + (A - 1) (2 \mathcal{S} + K^2_{ab} - K^2) - (5KA^{\prime} -  \frac {3} {2}\frac {A^{\prime 2}} {A} - 3A^{\prime\prime})
\label{a1}
\end{equation}
where $R$ is the Ricci scalar of the background metric (viz. the one with $A=1$), $R_{\Sigma}$ is the 3-dimensional Ricci scalar of $\sigma=$ constant surface, $\mathcal{S}\equiv R_{ab}n^an^b$,
$K_{ab}\equiv \nabla_an_b$ is the extrinsic curvature of $\sigma=$ constant surface and primes denote derivative with respect to $\sigma$. Obviously $R_{(q)} = R$ when $A=1$. 

We now need to take $\sigma\to0$ limit when $A\to L_P^2/\sigma^2$ \textit{diverges} and the limit needs to be taken with care. In particular, since $R$ (and all other curvature invariants) for the back ground ($A=1$) metric is assumed to be well-defined in the $\sigma\to0$ limit, we see that the $A^{-1}$ in first term in \eq{a1} kills the background Ricci scalar. \textit{Since this is the only term that survives when $L_P=0,A=1$, we can see the first signs of why the two limits do not commute.} To take the limit, we need an Taylor series expansion of various quantities. We start with the general result, valid in any spacetime:
\begin{equation}
\frac{1}{2}\nabla_a\nabla_b (\sigma^2) = g_{ab} - \frac{1}{3} \lambda^2 \mathcal{S}_{ab} + \frac{\lambda^3}{12} \nabla_n \mathcal{S}_{ab} - \frac{\lambda^4}{60} \lbrace \nabla_n \nabla_n \mathcal{S}_{ab} + \frac{4}{3}\mathcal{S}_{ia} \mathcal{S}^i_{b} \rbrace + 
{\mathcal O} (\lambda^5)
\label{expsigma}
\end{equation}
where $\lambda$ is the arc length of the geodesic connecting $p$ and $P$ which is numerically equal to $\sigma(p,P)$ and we use the notation $\nabla_n=n^j\nabla_j$. Further using the definition of $n_a$ in terms of $\sigma(x,x')$ (see second equation in \eq{n1}), it is easy to show that:
\begin{equation}
\lambda K_{ab} = \lambda \nabla_a n_b=\frac{1}{2}\nabla_a\nabla_b \sigma^2 - \epsilon n_a n_b
\end{equation}
Using \eq{expsigma} in this we can get the corresponding expansion for $K_{ab}$ as:
\begin{equation}
\lambda K_{ab} = h_{ab} - \frac {1} {3} \lambda^2 \mathcal{S}_{ab} + \frac {1} {12} \lambda^3\nabla_n \mathcal{S}_{ab} - \frac {1} {60}\lambda^4 F_{ab} + {\mathcal O} (\lambda^5)
\end{equation}
where
\begin{equation}
F_{ab} = \nabla_n \nabla_n \mathcal{S}_{ab} + (4/3) \mathcal{S}_{ia} \mathcal{S}^i_{b}
\end{equation}
Taking the trace we get:
\begin{equation}
\lambda K = D_1 - \frac {1} {3} \lambda^2 \mathcal{S} + \frac {1} {12} \lambda^3\nabla_n \mathcal{S} - \frac {1} {60}\lambda^4 F
\end{equation}
where we use the notation $D_k=D-k$ and
\begin{equation}
F = F_{ab} g^{ab} = \nabla^2_q \mathcal{S} + (4/3) \mathcal{S}_{ab} \mathcal{S}^{ab}
\end{equation}
Using these two results we can calculate the useful combination:
\begin{equation}
\lambda^2 (K^2_{ab} - K^2 ) = -D_1D_2 + \frac {2} {3} \lambda^2 D_2\mathcal{S} - \frac {1} {6} \lambda^3 D_2\bdot{\mathcal{S}} + \lambda^4 F_1 + \mathcal{O}(\lambda^5)
\end{equation}
where we have defined, for ease of notation, the quantity:
\begin{equation}
F_1\equiv \frac {1} {15} \l[\l( \frac {2D +1} {3} \mathcal{S}^2_{ab} + \frac {D_2} {2} \bddot{\mathcal{S}} - \frac {5} {3} \mathcal{S}^2 \r)\r]                                                                                                                                 \end{equation}
and use overdot to denote the operation of $n^i\nabla_i$.  
Finally, we compute the Taylor series expansion of $R_\Sigma$:
\begin{equation}
R_{\Sigma} = R - 2 \epsilon\mathcal{S} - \epsilon (K^2_{ab} - k^2) = R - 2 \epsilon\mathcal{S} -  \frac {\epsilon} {\lambda^2} \l( -D_1D_2 + \frac {2} {3} \lambda^2 D_2\mathcal{S} + \mathcal{O}(\lambda^3) \r)
\end{equation}
We are now in a position to evaluate all the terms which appear in the right hand side of \eq{a1}.
Note that, for our special case, we can set $\lambda=\sigma$ in the expansions. We will also set $D=4$.
Starting with the results
\begin{equation}
1 - A^{-1} = 1 - \frac {\sigma^2} {4^2} + \mathcal{O} (\sigma^4 / L_p^4) 
\end{equation}
and
\begin{equation}
R - R_{\Sigma} = \frac {10} {3} \mathcal{S} 
- \frac {6} {\sigma^2} + \mathcal{O} (\sigma) 
\end{equation}
we get the term $(1 - A^{-1}) (R - R_{\Sigma})$ in \eq{a1} to be:
\begin{equation}
(1 - A^{-1}) (R - R_{\Sigma}) = \frac {10} {3} \mathcal{S}  - \frac {6} {\sigma^2} + \frac {6}{L_p^2} + \mathcal{O} (\sigma) 
\label{a2}
\end{equation}
Next consider the term $(A - 1) ( 2 \mathcal{S} + K^2_{ab} - K^2)$ in \eq{a1}.
Using
\begin{equation}
K^2_{ab} - K^2 = - \frac {6} {\sigma^2} + \frac {4} {3} \mathcal{S} - \frac {1} {3}\sigma \bdot {\mathcal{S}} + \sigma^2F_1 + \mathcal{O} (\sigma^3) 
\end{equation}
we get:
\begin{equation}
(A - 1) ( 2 \mathcal{S} + K^2_{ab} - K^2) = - \frac {6L^2_p} {\sigma^4} + \frac {10} {3} \frac {L^2_p} {\sigma^2} \mathcal{S} - \frac {1} {3} \frac {L^2_p} {\sigma} \bdot{\mathcal {S}} + F_1 L^2_p + \mathcal{O} (\sigma)
\label{a3} 
\end{equation}
Similarly, using
\begin{equation}
K = \frac {3} {\sigma} - \frac {1} {3} \sigma \mathcal{S} + \frac {1} {12} \sigma^2 \bdot {\mathcal{S}} -\frac{1}{60}F\sigma^3 + \mathcal{O} (\sigma^4) 
\end{equation}
the last bunch of terms $[-5KA^{\prime} - (3/2) (A^{\prime 2}/A) - 3A^{\prime\prime}]$ in \eq{a1} evaluates to:
\begin{equation}
-5KA^{\prime} - (3/2) (A^{\prime 2}/A) - 3A^{\prime\prime} = \frac {6 L^2_p} {\sigma^{4}} - \frac {10} {3} \frac {L^2_p} {\sigma^2} \mathcal{S}  + \frac {5} {6}  \frac {L^2_p} {\sigma}  \bdot {\mathcal {S}} + \frac {6} {\sigma^2} -  \frac {6} {L^2_p} + F_3 L^2_p  +  \mathcal{O} (\sigma) 
\label{a4}
\end{equation}
where $F_3=-(1/6)F$. 
We now substitute these expressions in \eq{a2}, \eq{a3} and \eq{a4}  into \eq{a1} to obtain:
\begin{eqnarray}
{R_{(q)}} &=& A^{-1}R + \left\{ \frac {10} {3} \mathcal{S} - \frac {6} {\sigma^2} + 
\frac {6} {L^2_p} + \mathcal{O} (\sigma) \right\}  \nonumber\\
&+&\left\{-  \frac {6{L^2_p}} {\sigma^{4}} + \frac {10} {3} \frac {L^2_p} {\sigma^2} \mathcal{S} - \frac {1} {3} \frac {L^2_p} {\sigma} \bdot {\mathcal {S}} +  F_1L^2_p + \mathcal{O} (\sigma) \right\} \nonumber\\
&&+ \left\{ \frac {6{L^2_p}} {\sigma^{4}} - \frac {10} {3} \frac {L^2_p} {\sigma^2}
\mathcal{S} + \frac {6} {\sigma^2} - \frac {6} {L^2_p} + \frac {5} {6} \frac {L^2_p} {\sigma} \bdot {\mathcal {S}} + F_3 L^2_p + \mathcal{O} (\sigma)\right\}
\end{eqnarray}
Miraculously,  the terms diverging as $\sigma^{-4}$ and $\sigma^{-2}$ nicely cancel leaving only a 
$\sigma^{-1}$ divergence which we know how to regularize in the point-splitting approach. \textit{Further, the non-analytic terms in Planck length $1/L_P^2$ also cancel out.} Finally, the $A^{-1}$ factor kills the background Ricci scalar term, leading to the final result:
\begin{equation}
{R_{(q)}} \rightarrow \frac {10} {3} \mathcal{S} + (F_1 + F_3 ) L^2_p + \left[ \frac {\mathcal {\bdot S} L^2_p} {2 \sigma} \right]_{\sigma \to 0}
\end{equation}
The prefactor is indeed $(2/3)(D+1)=10/3$ for $D=4$ we are working in. Once the divergent expression is regularized and eliminated, the leading term is  proportional to 
$\mathcal{S}=R_{ab}n^an^b$ as advertised.

Exactly the same kind of analysis works in the most general background when we work in an arbitrary coordinate system. The above approach shows that taking $\sigma\to0$ first, as we should, picks out the correct combination of terms, finally leading to the entropy density.

\section{Aside on the geometrical structure of the modified curvature scalar} \label{app:geom-struc}

Eq. (\ref{eq:main-eq-rs}) gives the full expression for the modified Ricci scalar, and as is evident, its evaluation requires the knowledge of the geodesic interval biscalar. Nevertheless, it is instructive to rewrite 
it in a slightly different manner which highlights the elegant geometrical structure of the whole set-up, and might be helpful in further developments on purely geometric grounds. After a few algebraic manipulations, Eq. (\ref{eq:main-eq-rs}) can be written as
\begin{eqnarray}
\rmod(p;P,\lp) = A R - \l( A - A^{-1} \r) \l( R_{\Sigma} -  R^0_{\Sigma} \r) 
+
2 \epsilon (A-1) 
\l(\frac{D+1}{D-1}\r) 
\l( K - K^0 \r) K^0 
\nn \\
\end{eqnarray}
where $R^0_{\Sigma}={D_1 D_2}/{\sigma^2}$ and $K^0={D_1}/{\sqrt{\epsilon \sigma^2}}$ are the induced and extrinsic curvatures of the $\sigma^2=$const. surfaces in the {\it flat} spacetime (which are maximally symmetric spaces with positive or negative curvature). Note that the first two terms mimic the relationship between $g_{ab}$ and $\gm_{ab}$. It is worth investigating the geometrical structure of the above expression in detail, since it might hold the key to a generic study of the behavior of the modified curvature scalar near the spacetime singularities in terms of the focusing of geodesics. This work is currently in progress.

\section{The divergent term in Eq. (\ref{eq:mod-rs})} \label{app:div-term}
As mentioned in the paper, there is a divergent term in (\ref{eq:mod-rs}) which is given by
\begin{eqnarray}
\lim \limits_{\lambda \rightarrow 0^+}  +\frac{\lp^2}{2 \lambda} \bdot {\mathcal S}(p)
\end{eqnarray}
This term has an odd number of factors of $n^i$, since $\bdot\mathcal{S} = n^i \nabla_i \l( R_{jk} n^j n^k \r) = n^i n^j n^k \nabla_i R_{jk}$. (Recall that $n^i \nabla_i n^j=0$). Therefore, if one defines the coincidence limit by considering the limits $p,p' \rightarrow P$ where $p, p'$ are diametrically opposite, i.e.,
$$
\rmod_{\rm reg}(P,\lp) = \frac{1}{2} \Biggl( \underset{p \rightarrow P}{\lim} \rmod(p;P,\lp) + \underset{p' \rightarrow P}{\lim} \rmod(p';P,\lp) \Biggl)
$$
the divergent term will cancel out. (Other terms have even number of factors of $n^i$, so their contribution remains unchanged). This way of taking the limit was discussed by Christensen and Davies et al. (amongst others) in some older works related to the point-splitting regularization in curved spacetime \cite{christenson-worldf, DFCB}. We, therefore, obtain
$$\rmod_{\rm reg}(P,\lp) = \alpha \l[R_{ab} n^a n^b \r]_P - \frac{\lp^2}{15} \;
\l[ \frac{1}{3} \mathcal S_{ab} \mathcal S^{ab} - 6 \bddot {\mathcal S} + \frac{5}{3} \mathcal S^2 \r]_P
$$
Note that the coefficient of the $\bddot {\mathcal S}$ gets modified (see Eq. (\ref{eq:mod-rs})) since $\lambda^{-1} \bdot {\mathcal S}(p) = \lambda^{-1} \l[ \bdot {\mathcal S}(P) + \lambda \bddot {\mathcal S}(P) + \mathcal{O}(\lambda^2) \r]$ has a $\mathcal{O}(\lambda^0)$ contribution.



 \end{document}